\documentclass[aps,prb,reprint,superscriptaddress]{revtex4-2}
\usepackage{amsfonts}
\usepackage{amsmath}
\usepackage{amssymb}
\usepackage{graphicx}
\usepackage{xcolor}
\usepackage{epstopdf}
\usepackage{bm}%
\begin{document}
\title{Pressure-Driven Evolution of Electronic and Magnetic Correlations in Bilayer Nickelate La$_{3}$Ni$_{2}$O$_{7}$}


\author{Jian Zhou}
\altaffiliation{These authors contributed equally to this work}
\affiliation{National Key Laboratory of Surface Physics and Chemistry, Mianyang, 621908, China}

\author{Rui Song}
\altaffiliation{These authors contributed equally to this work}
\affiliation{National Key Laboratory of Surface Physics and Chemistry, Mianyang, 621908, China}

\author{Haiyan Lu}
\email{hyluphys@163.com}
\thanks{Corresponding author.}
\affiliation{National Key Laboratory of Surface Physics and Chemistry, Mianyang, 621908, China}

\begin{abstract}
The recent discovery of high-temperature superconductivity in pressurized bilayer La$_3$Ni$_2$O$_7$ has sparked intense research interest, yet the microscopic mechanism governing its pressure-dependent superconducting transition temperature ($T_c$) remains elusive. In this work, we investigate the electronic and magnetic correlations of La$_3$Ni$_2$O$_7$ under high pressure using a combination of density-functional theory (DFT), constrained random phase approximation (cRPA), and dynamical mean-field theory (DMFT). We find that while hydrostatic pressure enhances the interlayer hopping and the bare superexchange energy scale ($\sim4t^2/U$), it simultaneously drives the system toward a more itinerant regime by reducing the relative correlation strength ($U/W$). Crucially, our results reveal a distinct orbital-selective evolution: the Ni $d_{x^2-y^2}$ states become increasingly itinerant, whereas the Ni $d_{z^2}$ orbitals retain a more localized character. This pressure-induced itinerancy significantly enhances the hybridization between the two, leading to a dramatic amplification of the Kondo-like screening of the local $d_{z^2}$ moments by the itinerant $d_{x^2-y^2}$ electrons. Consequently, the effective magnetic exchange coupling ($J_{\text{eff}}$), which serves as the pairing glue, is suppressed in the high-pressure regime. Our findings suggest that the monotonic decrease of $T_c$ at high pressures is driven by the dominance of Kondo screening over superexchange interactions, providing a coherent microscopic explanation for the dome-shaped superconducting phase diagram in La$_3$Ni$_2$O$_7$.
\end{abstract}

\maketitle

\section{Introduction}
Nickelates have emerged as the third family of high-temperature superconductors, following the cuprates and iron-based superconductors~\cite{li2019superconductivity,sun2023signatures,chen2026nickelate_review}. However, unlike these two well-established families, achieving a superconducting transition temperature ($T_{c}$) above the liquid nitrogen temperature in bulk nickelates requires high-pressure conditions~\cite{hou2023emergence,wang2024pressure,li2025identification}. This unique prerequisite underscores the critical importance of investigating the pressure-driven tuning of superconducting properties in nickelates~\cite{hou2023emergence,li2025identification}. Furthermore, the current evolution of $T_{c}$ with pressure, particularly in the bilayer La$_3$Ni$_2$O$_7$ system, exhibits a phase diagram that differs obviously from the canonical phase diagrams of cuprates and iron-based superconductors~\cite{hou2023emergence,zhang2024zero,li2025identification,keimer2015cuprates,paglione2010iron}. This striking discrepancy naturally raises fundamental questions: Does the mechanism of high-temperature superconductivity in nickelates share the same origin as that in cuprates or iron-based superconductors? Can the theoretical frameworks developed for cuprate and iron-based superconductors be directly transplanted to understand nickelates~\cite{luo2023bilayer,zhang2024structural,qin2023singlets,luo2024tj}?
\par Additionally, previous computational studies on nickelate superconductors have typically relied on empirical values for crucial parameters, such as the Hubbard $U$ and Hund's coupling $J_H$, applying them universally across different pressure conditions~\cite{christiansson2023correlated,ouyang2024hund,liu2025orbital}. The explicit pressure dependence of these core parameters, however, remains poorly understood and has yet to be systematically clarified~\cite{aryasetiawan2004frequency,aryasetiawan2006calculations,verraes2026structural}. Moreover, in the microscopic theoretical study of superconducting mechanisms, there has long been a debate between the weak-coupling itinerant-electron picture and the strong-coupling local-moment picture~\cite{luo2023bilayer,zhang2024structural,qin2023singlets,wu2024superexchange,luo2024tj,pan2026review}. While both approaches have demonstrated unique advantages in explaining specific phenomena, they also suffer from inherent shortcomings that contradict experimental observations~\cite{pan2026review,chen2026nickelate_review}. For instance, the weak-coupling itinerant-electron picture successfully explains the suppression of $T_{c}$~\cite{luo2023bilayer,zhang2024structural} under high pressure in the ``327'' system, where the presence of the $\gamma$ pocket at the Fermi surface plays a crucial role in determining the enhancement or suppression of $T_{c}$. However, some experimental studies indicate that the existence of the $\gamma$ pocket has little impact on superconductivity~\cite{wang2026electronic,ko2025ambient}. Conversely, the strong-correlation local-moment picture argues that the $\gamma$ pocket is not essential, but within this framework, pressure is generally expected to enhance $T_{c}$~\cite{qin2023singlets,luo2024tj,pan2026review}. In light of these contradictions, we believe that elucidating the evolution of the localized and itinerant nature of electrons under pressure constitutes a highly valuable research direction~\cite{yang2024orbital,christiansson2023correlated,ouyang2024hund, liu2025orbital,lu2024interplay,pan2026review}.
\par In this work, we first employ the constrained random phase approximation (cRPA) to determine the correlation parameters of the Ni $d$ orbitals in the bilayer ``327'' system under various pressures~\cite{aryasetiawan2004frequency,aryasetiawan2006calculations}. This allows us to obtain a more precise description of the pressure-driven evolution of its electronic structure. We find that while hydrostatic pressure enhances the interlayer hopping and the bare superexchange energy scale, it simultaneously drives the system toward a more itinerant regime by reducing the relative correlation strength. Subsequently, through dynamical mean-field theory (DMFT) calculations, we find a distinct orbital-selective evolution: the Ni $d_{x^2-y^2}$ states become increasingly itinerant, whereas the Ni $d_{z^2}$ orbitals retain a more localized character. This pressure-induced itinerancy significantly enhances the hybridization between the two, leading to a dramatic amplification of the Kondo-like screening of the local $d_{z^2}$ moments by the itinerant $d_{x^2-y^2}$ electrons, which is likely a crucial factor contributing to the suppression of $T_{c}$ under pressure.

\section{Methods}
\subsection{Constrained Random Phase Approximation (cRPA)}

The Coulomb interaction parameters were evaluated from first principles using the constrained random phase approximation (cRPA)~\cite{aryasetiawan2004frequency,aryasetiawan2006calculations}. In cRPA, the full electronic structure is downfolded onto an effective low-energy Hamiltonian spanned by the correlated orbitals:
\begin{equation}
H = \sum_{ij R}t_{ij}^{(R)}c_{i R}^{\dagger}c_{j R} + \sum_{ijkl R}U_{ijkl}c_{i R}^{\dagger}c_{j R}^{\dagger}c_{l R}c_{k R},
\end{equation}
where $t_{ij}^{(R)}$ is the hopping matrix element between Wannier orbitals $|w_i\rangle$ and $|w_j\rangle$ at lattice vector $R$, constructed as maximally localized Wannier functions from the Kohn-Sham eigenstates~\cite{marzari1997maximally}. The effective on-site interaction matrix element $U_{ijkl}$ is defined as
\begin{equation}
U_{ijkl} = \lim_{\omega\rightarrow 0}\iint dr~dr^{\prime}w_{i}^{*}(r)w_{j}^{*}(r^{\prime})U(r,r^{\prime},\omega)w_{l}(r)w_{k}(r^{\prime}).
\end{equation}

The partially screened Coulomb kernel $U$ is obtained from the constrained polarizability $\chi^r$ through
\begin{equation}
U^{-1} = V^{-1} - \chi^{r},
\end{equation}
where $V$ is the bare Coulomb kernel. The constrained polarizability is defined as $\chi^r=\chi-\chi^c$, where $\chi$ is the total independent-particle polarizability at the RPA level and $\chi^c$ is the polarizability within the correlated subspace~\cite{bohm1951collective,pines1952collective,bohm1953collective,gell1957correlation,nozieres1958correlation}. Excluding $\chi^c$ avoids double counting of screening processes later treated explicitly in the correlated model.

\par A key step in cRPA is defining $\chi^c$. This is straightforward for isolated correlated bands but nontrivial when the target bands are entangled with other states. Several schemes have been proposed. The disentanglement method isolates the correlated subspace by separate diagonalization~\cite{aryasetiawan2006calculations}, but it modifies the band structure and depends on the chosen Wannier energy window~\cite{miyake2009ab}. The weighted method preserves the band structure by weighting states according to their Wannier projection probabilities~\cite{csacsiouglu2011effective}, but neglects important non-diagonal polarizability terms~\cite{kaltak2015dmft}. In this work, the projector method based on the Kubo-Nakano formula is adopted because it preserves the physical band structure and retains non-diagonal contributions. This method introduces the projected Bloch function $|\bar{\psi}_{nk}\rangle$~\cite{kaltak2015dmft}:
\begin{equation}
|\bar{\psi}_{nk}\rangle = \sum_{m}P_{mn}^{(k)}|\psi_{mk}\rangle.
\end{equation}
Here, $|\psi_{nk}\rangle$ are Kohn-Sham eigenstates, and $P_{mn}^{(k)}$ projects them onto the correlated subspace using the unitary matrices from the Wannier construction. The correlated polarizability is then calculated as~\cite{kaltak2015dmft}
\begin{equation}
\begin{aligned}
&\chi^c(\mathbf{r}, \mathbf{r}'; \omega) = \sum_{n,n'} \bar{\psi}_n(\mathbf{r})\bar{\psi}_{n'}^*(\mathbf{r})\bar{\psi}_n^*(\mathbf{r}')\bar{\psi}_{n'}(\mathbf{r}') \\
&\quad \times \left( \frac{1}{\omega + (E_n - E_{n'}) + i\eta} - \frac{1}{\omega - (E_n - E_{n'}) - i\eta} \right).
\end{aligned}
\end{equation}

\par After obtaining $U_{ijkl}(\omega)$, the static Hubbard-Kanamori parameters, including the intraorbital interaction $U$, interorbital interaction $U^\prime$, and Hund's coupling $J$, are extracted at $\omega=0$. For a correlated subspace with $N$ states, they are defined as~\cite{kanamori1963electron,vaugier2012hubbard}
\begin{equation}
U = \frac{1}{N}\sum_{i}^{N}U_{iiii},
\end{equation}
\begin{equation}
U^{\prime} = \frac{1}{N(N-1)}\sum_{i\ne j}^{N}U_{ijji},
\end{equation}
\begin{equation}
J = \frac{1}{N(N-1)}\sum_{i\ne j}^{N}U_{ijij}.
\end{equation}

\subsection{DFT+DMFT calculation}

The electronic structure of La$_3$Ni$_2$O$_7$ was investigated using density-functional theory (DFT) combined with single-site dynamical mean-field theory (DMFT) to treat the electronic correlations of the Ni 3$d$ orbitals~\cite{georges1996dynamical,kotliar2006publication}. The crystal structures optimized using VASP were adopted in the calculations~\cite{kresse1993ab,kresse1994ab,kresse1994norm,kresse1996efficiency,kresse1996efficient,kresse1999ultrasoft}. The Perdew-Burke-Ernzerhof generalized gradient approximation was used for the exchange-correlation potential~\cite{perdew1996generalized}. Brillouin zone integration was performed on a $12 \times 12 \times 12$ $k$-mesh, corresponding to 159 inequivalent $k$-points in the first irreducible Brillouin zone. Self-consistent calculations were performed using the full-potential linearized augmented plane-wave code WIEN2k~\cite{blaha2001wien2k}. The cutoff parameter was set to $R_{\text{MT}}K_{\text{max}} = 7.0$, and the muffin-tin radii of La and Ni were fixed to 2.34 and 1.90 a.u., respectively. 

Fully charge-self-consistent DFT+DMFT calculations were performed using the EDMFT package~\cite{haule2010dynamical}. Convergence was typically reached within approximately 30 DFT+DMFT cycles. The convergence criteria for charge and total energy were $10^{-5}$~e and $10^{-5}$~Ry, respectively. Each cycle contained one DMFT step and up to 100 DFT iterations. The calculations were performed at $\beta = 145.0$, corresponding to approximately 80~K, with the system constrained to be paramagnetic. The Ni 3$d$ orbitals were treated as correlated. The Coulomb interaction matrix was constructed from Slater integrals~\cite{fujiwara1999spin}, using the cRPA-derived effective Coulomb interaction $U$ and Hund's exchange parameter $J_H$~\cite{aryasetiawan2004frequency,aryasetiawan2006calculations}. The exact double-counting scheme was used for the self-energy. Neglecting off-diagonal hybridization function elements, the resulting Anderson impurity models were solved via the hybridization-expansion continuous-time quantum Monte Carlo (CT-HYB) solver, with $1 \times 10^{8}$ Monte Carlo sweeps performed for each impurity model.
The Matsubara self-energy was analytically continued using the maximum entropy method (MEM)~\cite{jarrell1996maximum} when the convergence criteria are reached. The resulting real-frequency self-energy was used to calculate the momentum-resolved spectral function $A(\mathbf{k},\omega)$ and the density of states $A(\omega)$.

\subsection{Calculation of magnetic exchange interactions}

Magnetic exchange interactions in La$_3$Ni$_2$O$_7$ were calculated by combining \textsc{ABACUS}~\cite{chen2010systematically,li2016large} first-principles calculations with the \textsc{TB2J} package~\cite{he2021tb2j}. Spin-polarized DFT calculations were first performed in \textsc{ABACUS}, and the resulting localized-orbital Hamiltonian and overlap matrices were used as input for \textsc{TB2J}. The magnetic interactions were then evaluated from the corresponding tight-binding-like Hamiltonian in the localized atomic-orbital basis.

The magnetic interactions were mapped onto an effective extended Heisenberg Hamiltonian,
\begin{equation}
\begin{aligned}
E =&-\sum_i K_i(\vec{S}_i\cdot\vec{r}_i)^2 \\
&-\sum_{i\neq j}
\left[
J^{\mathrm{iso}}_{ij}\vec{S}_i\cdot\vec{S}_j
+
\vec{S}_i \mathbf{J}^{\mathrm{ani}}_{ij}\vec{S}_j
+
\mathbf{D}_{ij}\cdot
\left(
\vec{S}_i\times\vec{S}_j
\right)
\right].
\end{aligned}
\end{equation}
Here, $K_i$ denotes the single-ion anisotropy, $J^{\mathrm{iso}}_{ij}$ is the isotropic exchange interaction, $\mathbf{J}^{\mathrm{ani}}_{ij}$ is the symmetric anisotropic exchange tensor, and $\mathbf{D}_{ij}$ is the Dzyaloshinskii--Moriya interaction vector. The spin vectors $\vec{S}_i$ are normalized to unity, so the exchange parameters are expressed in units of energy. Under this convention, positive and negative $J^{\mathrm{iso}}_{ij}$ correspond to ferromagnetic and antiferromagnetic coupling, respectively.

The exchange parameters were evaluated within the magnetic force theorem using the local rigid spin-rotation approximation implemented in \textsc{TB2J}. The single-particle Green's function was constructed from the Hamiltonian $H(\mathbf{k})$ and overlap matrix $S(\mathbf{k})$ as~\cite{he2021tb2j}
\begin{equation}
G(\mathbf{k},\epsilon)
=
\left[
\epsilon S(\mathbf{k}) - H(\mathbf{k})
\right]^{-1}.
\end{equation}
The real-space Green's function was obtained by Fourier transformation and used to evaluate pairwise magnetic interactions between Ni sites.

For collinear calculations without spin--orbit coupling, the isotropic exchange interaction was calculated using the Liechtenstein--Katsnelson--Antropov--Gubanov Green's-function expression~\cite{liechtenstein1987local},
\begin{equation}
J^{\mathrm{iso}}_{ij}
=
-\frac{1}{4\pi}
\int_{-\infty}^{E_F}
\mathrm{Im}\,
\mathrm{Tr}
\left[
\Delta_i
G^{\uparrow}_{ij}
\Delta_j
G^{\downarrow}_{ji}
\right]
d\epsilon,
\end{equation}
where $\Delta_i$ is the on-site exchange splitting on magnetic site $i$, $G^{\uparrow}_{ij}$ and $G^{\downarrow}_{ji}$ are spin-resolved real-space Green's functions, and the trace is taken over local orbitals. The energy integration was performed by contour integration up to the Fermi level $E_F$. The exchange parameters can be further decomposed into orbital-pair contributions~\cite{liechtenstein1987local}:
\begin{equation}
    J^{\mathrm{iso}}_{im,jm'} = -\frac{1}{4\pi} \mathrm{Im} \int_{-\infty}^{E_F} \Delta_{im} G^{\uparrow}_{im,jm'} \Delta_{jm'} G^{\downarrow}_{jm',im} d\epsilon.
\end{equation}

Ni atoms were selected as the magnetic centers. Convergence with respect to the $k$-point mesh, localized-orbital basis, and interaction cutoff was carefully checked. The final exchange constants were grouped according to inequivalent Ni--Ni exchange paths. The nearest-neighbor interlayer exchange interaction within each NiO$_2$ bilayer was selected for the subsequent analysis.

\section{Results}
\subsection{Results of Electronic Correlation Calculations}

In the present calculations, the optimized tetragonal $I4/mmm$ structure was adopted at all investigated pressures, including 14.1~GPa.

\begin{figure}[htbp]
    \centering
    \includegraphics[width=0.95\columnwidth]{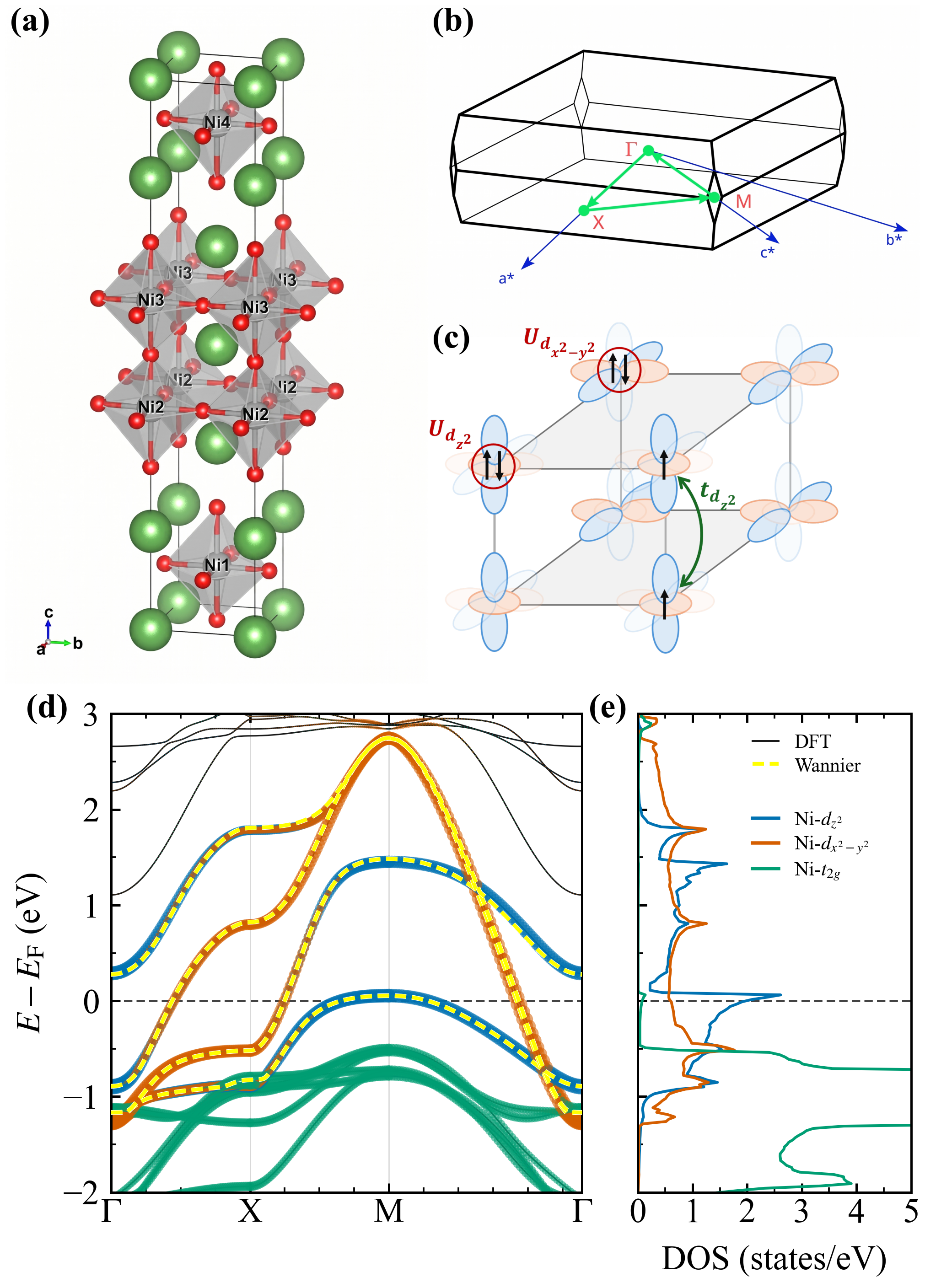}
    \caption{Representative crystal and orbital-resolved electronic structures of La$_3$Ni$_2$O$_7$ at 14.1~GPa. (a) Conventional cell of the $I4/mmm$ phase of La$_3$Ni$_2$O$_7$, visualized using VESTA~\cite{momma2011vesta}. (b) Brillouin zone and high-symmetry path generated using XCrySDen~\cite{kokalj1999xcrysden}. (c) Schematic illustration of the bilayer two-orbital model. (d) DFT band structure projected onto the Ni $d_{z^2}$, Ni $d_{x^2-y^2}$, and Ni $t_{2g}$ orbitals. The yellow dashed lines denote the Wannier-interpolated bands. (e) Orbital-projected density of states. The Fermi level is set to zero.}
    \label{structure}
\end{figure}

Recent high-pressure measurements on bilayer Ruddlesden-Popper La$_3$Ni$_2$O$_7$ single crystals have revealed pressure-induced structural transitions that are closely related to the emergence of superconductivity~\cite{li2025identification}. At low pressure, La$_3$Ni$_2$O$_7$ crystallizes in the orthorhombic $Amam$ structure. Upon compression, it transforms into a high-pressure orthorhombic $Fmmm$ phase near $14$~GPa, followed by a tetragonal $I4/mmm$ phase at higher pressure. Superconductivity appears in the high-pressure structural regime, with the onset transition temperature reaching approximately $83$~K near $18$~GPa and being progressively suppressed at higher pressures~\cite{sun2023signatures}. This pressure dependence indicates that superconductivity in La$_3$Ni$_2$O$_7$ is strongly coupled to the pressure-driven reconstruction of the bilayer NiO$_6$ framework.

To establish the structural and orbital framework underlying the subsequent correlation analysis, we first examine the crystal structure and low-energy electronic states of La$_3$Ni$_2$O$_7$ in the optimized $I4/mmm$ phase. As shown in Fig.~\ref{structure}(a), each Ni ion is octahedrally coordinated by six surrounding O ligands, forming a NiO$_6$ octahedron. The nominal Ni valence is Ni$^{2.5+}$, corresponding to an average $3d^{7.5}$ electron configuration. The octahedral crystal field splits the Ni $3d$ states into a lower-energy, nearly filled $t_{2g}$ manifold and a higher-energy $e_g$ manifold. The partially occupied $e_g$ orbitals, mainly Ni $3d_{x^2-y^2}$ and $3d_{z^2}$, dominate the low-energy electronic structure and therefore play a central role in the electronic, magnetic, and superconducting properties of La$_3$Ni$_2$O$_7$. The widely used bilayer two-orbital model in theoretical studies is illustrated in Fig.~\ref{structure}(c). In this bilayer square lattice, each site hosts two orbitals. The interlayer hybridization between the Ni $d_{z^2}$ orbitals is mediated by the apical O $p_z$ orbital, which leads to the splitting of the Ni $d_{z^2}$ bands into bonding and antibonding branches. In the tight-binding model, this interlayer hopping is described by the parameter $t_{d_{z^2}}$.

We next identify the orbital degrees of freedom that constitute the low-energy correlated subspace within the established bilayer geometry. The orbital-resolved band structure and density of states of La$_3$Ni$_2$O$_7$ at 14.1~GPa are presented in Figs.~\ref{structure}(d) and~\ref{structure}(e), respectively. Corresponding results at 30, 45, 60, 75, and 90~GPa are provided in Appendix~\ref{sec:appendix}. Around the Fermi level, the electronic states are mainly derived from the Ni $e_g$ orbitals, namely $d_{z^2}$ and $d_{x^2-y^2}$, while the Ni $t_{2g}$ states are located mostly below the Fermi level. The bonding $d_{z^2}$ branch is nearly occupied and exhibits a relatively narrow bandwidth, implying a more localized orbital character. In contrast, the Ni $d_{x^2-y^2}$ bands show stronger in-plane dispersion, reflecting the itinerant nature within the NiO$_2$ layers.

To verify that this low-energy orbital manifold can be represented reliably in the subsequent downfolding calculations, we compare the DFT bands with the corresponding Wannier interpolation. The yellow dashed curves in Fig.~\ref{structure}(d) denote the Wannier-interpolated bands obtained from maximally localized Wannier functions constructed for the Ni $d_{z^2}$ and $d_{x^2-y^2}$ orbitals. The excellent agreement between the Wannier-interpolated bands and the DFT $e_g$ bands demonstrates that the low-energy target subspace is accurately represented. This is essential for the subsequent constrained random-phase approximation calculation, because the screened interaction parameters depend sensitively on the definition of the correlated subspace, the orbital character of the low-energy bands, and the separation between target and screening channels. Therefore, the well-constructed Wannier orbitals provide a reliable basis for extracting effective Coulomb and Hund interaction parameters in pressurized La$_3$Ni$_2$O$_7$.

\begin{figure*}[htbp]
    \centering
    \includegraphics[width=0.95\textwidth]{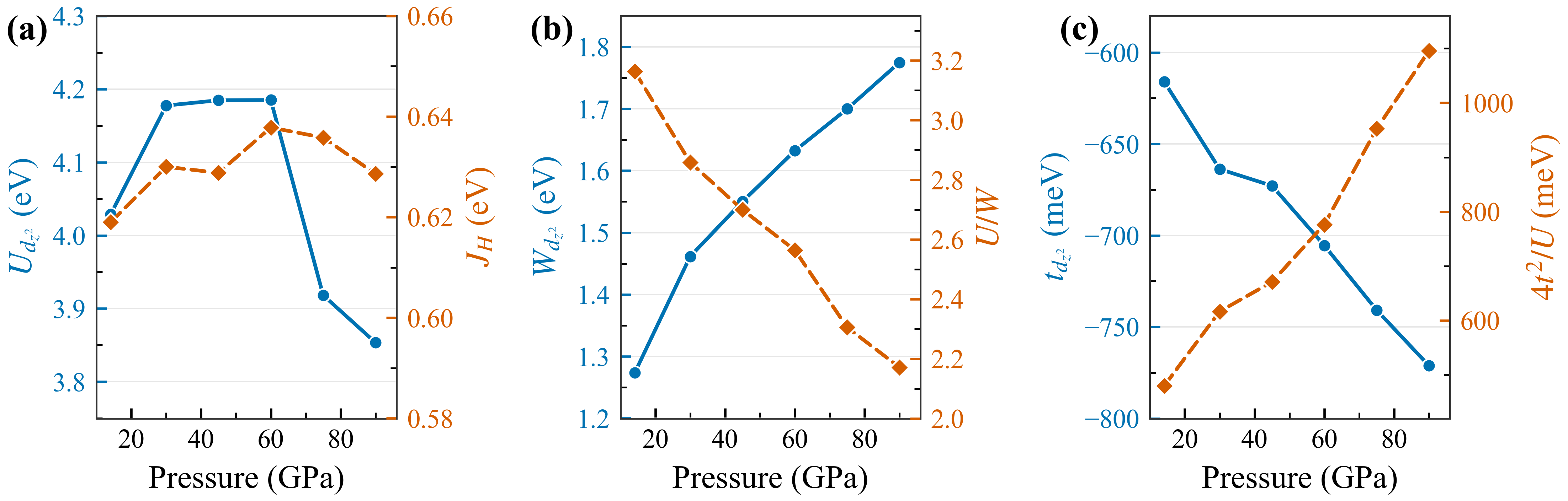}
    \caption{Pressure evolution of the cRPA interaction parameters and related electronic energy scales in La$_3$Ni$_2$O$_7$.
    (a) Screened on-site Coulomb interaction $U_{d_{z^2}}$ for the Ni $d_{z^2}$ orbital and Hund's exchange interaction $J_H$ between the Ni $d_{z^2}$ and $d_{x^2-y^2}$ orbitals.
    (b) Bandwidth $W_{d_{z^2}}$ of the Ni $d_{z^2}$ bonding states and the corresponding relative correlation strength $U_{d_{z^2}}/W_{d_{z^2}}$.
    (c) Interlayer hopping matrix element $t_{d_{z^2}}$ between two Ni $d_{z^2}$ orbitals sharing the same inner apical O atom, together with the estimated superexchange energy scale $4t_{d_{z^2}}^2/U_{d_{z^2}}$.}
    \label{crpa}
\end{figure*}

To quantify how pressure modifies the balance between local Coulomb
interactions, electronic itinerancy, and interlayer magnetic coupling,
we next evaluate the effective interaction and hopping parameters
within the Wannier subspace. Based on the DFT band structures and
well-converged Wannier functions constructed from the Ni $d_{z^2}$ and
$d_{x^2-y^2}$ orbitals, we downfolded the electronic structure onto the
low-energy correlated subspace and evaluated the effective interaction
parameters using the constrained random-phase approximation.
Figure~\ref{crpa}(a) shows the pressure dependence of the screened
on-site Coulomb interaction $U_{d_{z^2}}$ and Hund's exchange
interaction $J_H$. The screened $U_{d_{z^2}}$ displays a nonmonotonic pressure
dependence: it increases to approximately $4.18$~eV in the intermediate-pressure
range of 30$\sim$60~GPa and then decreases to about $3.85$~eV at 90~GPa.
This trend reflects a competition between pressure-enhanced local correlations
and enhanced electronic screening. At low and intermediate pressures, lattice
compression modifies the bilayer NiO$_6$ geometry and strengthens the
hybridization of the $d_{z^2}$-derived states, which can increase the local
Coulomb matrix element in the Wannier basis. At higher pressure, further band
broadening and stronger Ni $3d$-ligand hybridization enhance residual
polarization channels outside the correlated subspace, leading to stronger
screening and hence a reduced effective $U_{d_{z^2}}$. The maximum of
$U_{d_{z^2}}$ therefore marks the crossover between these two competing effects.
In contrast, $J_H$ remains nearly pressure independent, staying close to
$0.63$~eV over the whole pressure range. This weak pressure dependence is
expected because Hund's exchange is mainly governed by the intra-atomic
exchange within the Ni $3d$ shell and is less sensitive to charge screening than
the monopole Coulomb interaction $U$. These results indicate that the
low-energy physics of pressurized La$_3$Ni$_2$O$_7$ remains governed by a
robust Hund-coupled Ni $e_g$ manifold, while the effective Coulomb interaction
is tuned by the pressure-dependent screening environment.

Because the absolute value of $U_{d_{z^2}}$ alone does not determine
the correlation regime, we further compare it with the relevant
kinetic-energy scale. The evolution of the $d_{z^2}$ bonding-state
bandwidth $W_{d_{z^2}}$ and the relative correlation strength
$U_{d_{z^2}}/W_{d_{z^2}}$ is summarized in
Fig.~\ref{crpa}(b). With
increasing pressure, lattice compression enhances the spatial overlap between
neighboring Wannier orbitals, leading to a monotonic increase of
$W_{d_{z^2}}$ from approximately $1.27$~eV at 14.1~GPa to $1.77$~eV at
90~GPa. As a result, although the screened Coulomb interaction remains sizable,
the ratio $U/W$ decreases continuously with pressure. This trend indicates that
hydrostatic pressure drives the Ni $d_{z^2}$ bonding states toward a more
itinerant regime and weakens the relative strength of electronic correlations
in the highly compressed structures.

To determine whether the pressure-enhanced interlayer overlap also
strengthens the bare magnetic coupling between the two NiO$_2$ layers,
we finally examine the interlayer hopping and its associated
superexchange scale. Figure~\ref{crpa}(c) shows the pressure evolution
of the interlayer $d_{z^2}$ hopping matrix element. The two Ni $d_{z^2}$ orbitals in a bilayer
form strong $\sigma$ bonds through the shared inner apical O atom. Upon
compression, the magnitude of the interlayer hopping $|t_{d_{z^2}}|$ increases
substantially, reflecting the enhanced Ni--O--Ni orbital overlap along the
$c$ direction. Since this hopping amplitude grows more rapidly than the
screened interaction changes, the estimated superexchange scale $4t^2/U$
increases monotonically and reaches a value on the order of $1$~eV at high pressure.

This result has direct implications for the interpretation of superconductivity
in pressurized La$_3$Ni$_2$O$_7$. In a purely local-moment strong-coupling
picture, such as a conventional $t$-$J$-type scenario, an increasing
superexchange scale would generally be expected to strengthen magnetic pairing
interactions and thereby enhance the superconducting transition temperature.
However, experiments show that $T_c$ is suppressed when pressure is increased
beyond the optimal region. The simultaneous increase of $4t^2/U$ and decrease
of $U/W$ therefore suggest that the pressure dependence of superconductivity
cannot be explained solely by the growth of a local superexchange energy scale.
Instead, the results are more naturally consistent with a picture in which
itinerancy and correlation strength must be considered together: while pressure
enhances the bare interlayer hopping and superexchange scale, it also weakens
the relative electronic correlations and may suppress the spin fluctuations
needed to mediate Cooper pairing in the high-pressure regime.

\subsection{DMFT Results: Orbital Hybridization and Localized-Itinerant Characteristics}
\begin{figure*}[htbp]
    \centering
    \includegraphics[width=0.95\textwidth]{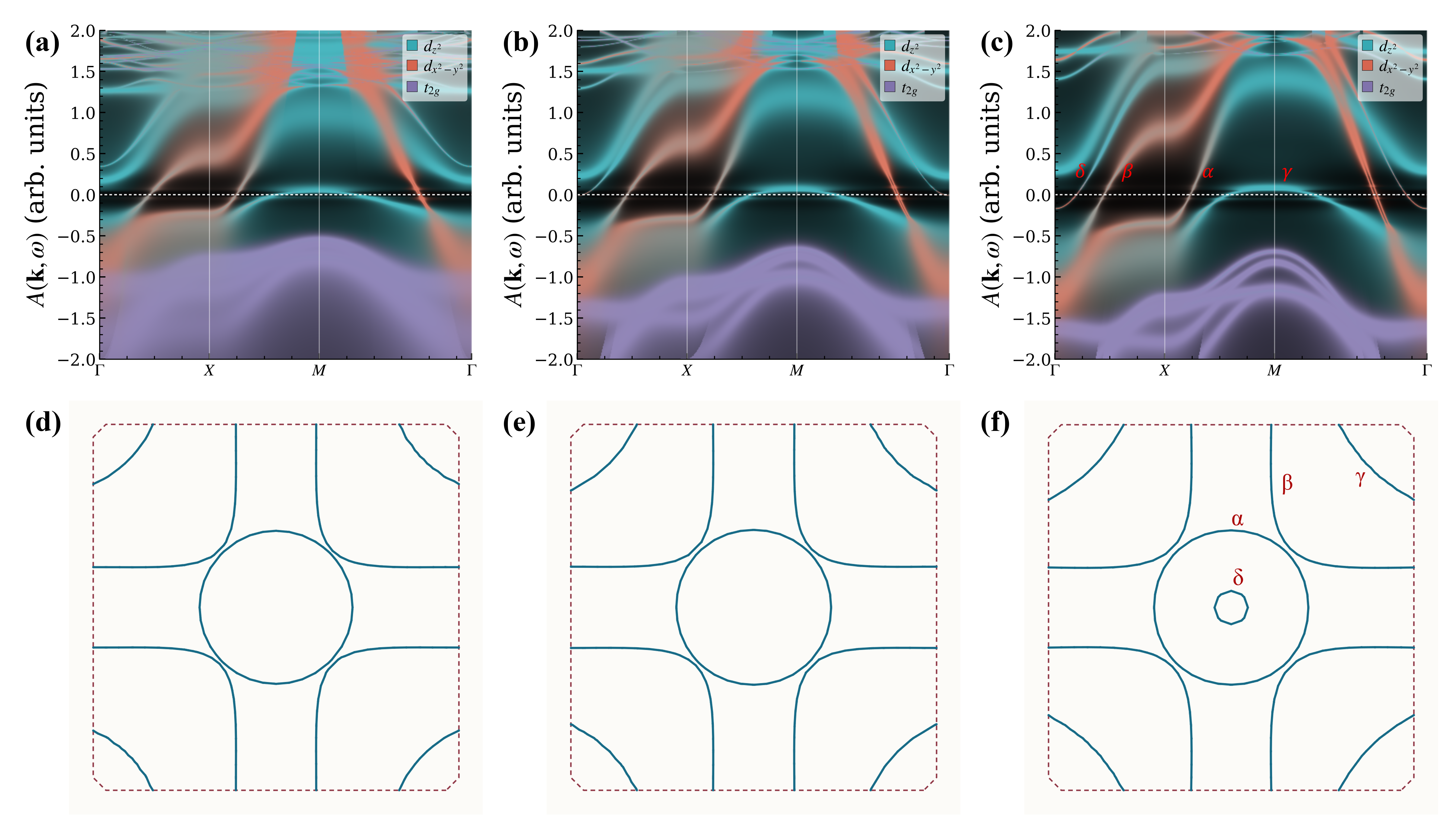}
    \caption{DFT+DMFT electronic structure of La$_3$Ni$_2$O$_7$ at $T = 80$~K using the pressure-dependent $U$ and $J_H$ parameters obtained from cRPA.
    (a)--(c) Momentum-resolved spectral functions at 14.1, 45, and 75~GPa, respectively.
    (d)--(f) Corresponding Fermi-surface sections at the same pressures.}
    \label{fig:band_dmft}
\end{figure*}

The cRPA results reveal competing pressure trends in the interaction,
bandwidth, and superexchange scales; it is therefore necessary to
determine how these trends reshape the correlated low-energy electronic
structure. Figure~\ref{fig:band_dmft} presents the DFT+DMFT
momentum-resolved spectral functions and corresponding Fermi-surface
sections calculated at $T = 80$~K. The Coulomb interaction $U$ and Hund's coupling $J_H$ used in
the DMFT calculations are taken from the pressure-dependent cRPA
results, providing a consistent connection between the downfolded
interaction parameters and the correlated electronic structure.
Compared with the DFT bands, the low-energy quasiparticle bands are
substantially renormalized. The Ni $d_{z^2}$-derived band near the
$M$ point becomes flatter and narrower, indicating stronger
orbital-dependent correlation effects and enhanced localized character.
By contrast, the Ni $d_{x^2-y^2}$-derived bands remain more dispersive,
consistent with their more itinerant in-plane nature in the NiO$_2$
layers. The nearly occupied Ni $t_{2g}$ states remain mainly below the
Fermi level and contribute weakly to the low-energy quasiparticle
structure, so the states near $E_F$ are dominated by the two Ni $e_g$
orbitals.

Having identified the orbital-dependent band renormalization, we next
examine whether pressure also induces qualitative changes in the
correlated Fermi-surface topology. The Fermi-surface sections in
Figs.~\ref{fig:band_dmft}(d)--(f) show the pressure evolution of the
correlated quasiparticle states.
The dominant $\alpha$ and $\beta$ sheets remain mainly derived from the
Ni $e_g$ bands, but their size and shape evolve continuously with
pressure. Meanwhile, the Ni $d_{z^2}$-derived $\gamma$ pocket becomes
more visible as the corresponding quasiparticle band approaches and
crosses the Fermi level. At 75~GPa, a small additional $\delta$ pocket
appears inside the central sheet, indicating a pressure-induced change
in the Fermi-surface topology. These results show that pressure not only deforms the existing $\alpha$ and $\beta$ sheets but also
promotes the emergence of new low-energy pockets through the combined
effects of lattice compression, orbital-selective correlation, and
pressure-dependent interaction strengths.

\begin{figure}[htbp]
    \centering
    \includegraphics[width=0.95\columnwidth]{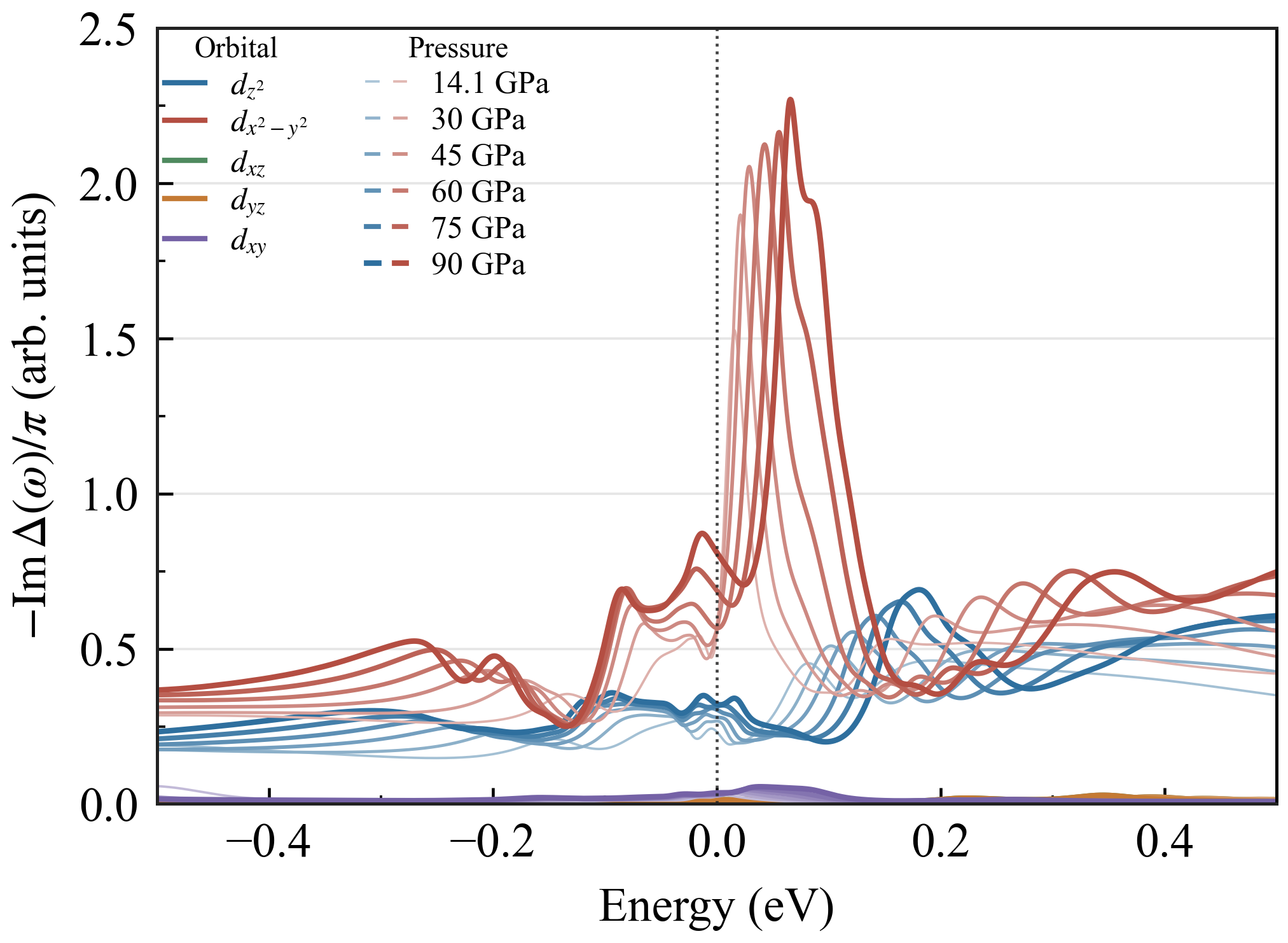}
    \caption{Orbital-resolved hybridization function $-\mathrm{Im}\Delta(\omega)/\pi$ of the Ni $d$ orbitals in La$_3$Ni$_2$O$_7$ under pressure from 14.1 to 90~GPa. The vertical dashed line marks the Fermi level.}
    \label{fig:dlt}
\end{figure}

Although the spectral functions indicate the orbital-dependent renormalization, a more direct characterization of the coupling between each local orbital and conduction bands is provided by the DMFT hybridization function. Figure~\ref{fig:dlt} shows the orbital-resolved quantity $-\mathrm{Im}\Delta(\omega)/\pi$ obtained from the DFT+DMFT calculations. In the DMFT impurity problem, the hybridization function $\Delta(\omega)$ describes how a local correlated Ni $d$ orbital is coupled to the surrounding electronic bath. It enters the impurity Green's function as
\[
G_{\mathrm{imp}}^{-1}(\omega)
=
\omega+\mu-\epsilon_{\mathrm{imp}}
-\Delta(\omega)-\Sigma(\omega),
\]
where $\epsilon_{\mathrm{imp}}$ is the local orbital energy and $\Sigma(\omega)$ is the local many-body self-energy. Therefore, $\Delta(\omega)$ encodes the one-particle coupling between the selected local orbital and all itinerant degrees of freedom outside the impurity, including ligand O-$2p$ states, neighboring Ni orbitals, and other bands included in the bath.

The quantity $-\mathrm{Im}\Delta(\omega)/\pi$ can be regarded as the bath spectral density seen by a given local orbital. Equivalently, it measures the energy-resolved strength of the effective hybridization channel between that orbital and the conduction bands. A large value of $-\mathrm{Im}\Delta(\omega)/\pi$ near the Fermi level indicates strong coupling to conduction bands, stronger quasiparticle broadening, and a more delocalized orbital character. In contrast, a small value near the Fermi level implies that the orbital is more weakly connected to the bath and therefore retains localized state. It should be noted that $-\mathrm{Im}\Delta(\omega)/\pi$ is not simply a single hopping parameter; rather, it reflects the combined effect of hopping matrix elements, orbital hybridization, and the density of available bath states at energy $\omega$.

With this interpretation, the pressure evolution of $-\mathrm{Im}\Delta(\omega)/\pi$ can be used to determine which orbital provides the increasingly itinerant screening channel. As pressure increases from 14.1 to 90~GPa, the hybridization spectrum of the Ni $d_{x^2-y^2}$ orbital is strongly enhanced near the Fermi level. The main peak on the unoccupied side, located slightly above $\omega=0$, increases from approximately $1.5$ at 14.1~GPa to above $2.3$ at 90~GPa. This enhancement indicates that compression strengthens the coupling between the $d_{x^2-y^2}$ orbital and the itinerant bath, consistent with the increased in-plane dispersion and enlarged bandwidth discussed above. The pressure evolution of the $d_{x^2-y^2}$ hybridization therefore reflects an increasingly itinerant in-plane electronic channel.

To assess the orbital selectivity of this pressure response, we compare the strongly enhanced $d_{x^2-y^2}$ channel with the corresponding $d_{z^2}$ hybridization. The Ni $d_{z^2}$ hybridization evolves more moderately over the same energy window. Although the $d_{z^2}$ orbital is crucial for the interlayer bonding channel through the inner apical oxygen, its low-energy hybridization remains smoother and less sharply enhanced than that of $d_{x^2-y^2}$. This behavior is consistent with the more localized character of the nearly occupied $d_{z^2}$ bonding state. The $t_{2g}$ components remain weak in the low-energy window, in agreement with their nearly filled character and minor role in the correlated low-energy physics. Overall, the hybridization function supports the orbital-selective behavior where pressure mainly enhances the itinerancy of the $d_{x^2-y^2}$ orbital, while the $d_{z^2}$ bonding state remains comparatively more localized.

\begin{figure*}[htbp]
    \centering
    \includegraphics[width=0.8\textwidth]{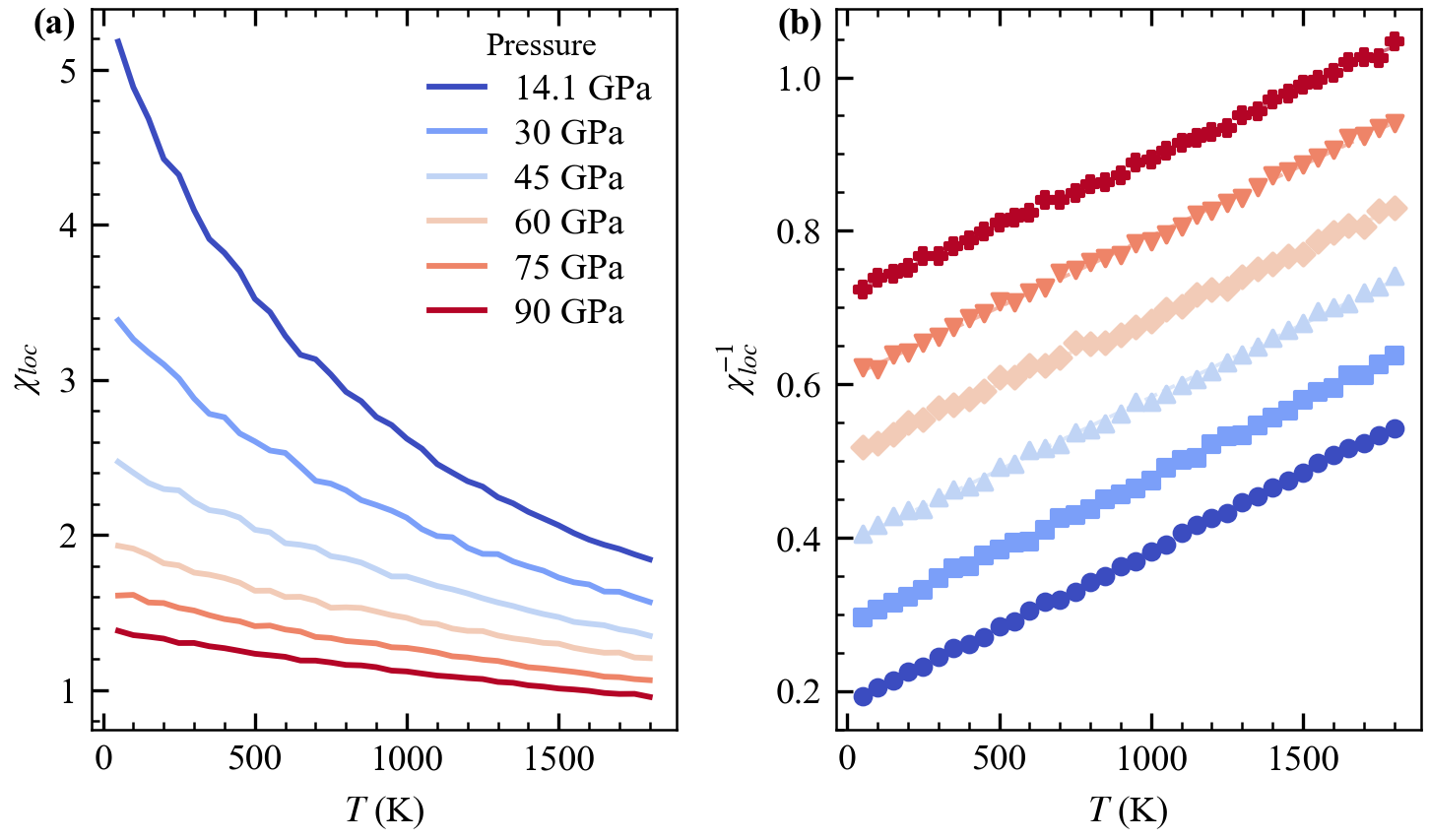}
    \caption{Temperature dependence of the local spin response in La$_3$Ni$_2$O$_7$ under pressure.
    (a) Static local spin susceptibility $\chi_{\mathrm{loc}}$ as a function of temperature.
    (b) Inverse local spin susceptibility $\chi_{\mathrm{loc}}^{-1}$ and the corresponding Curie-Weiss fits. The dashed lines denote fits over the temperature range 50$\sim$1800 K.}
    \label{fig:susc}
\end{figure*}

The pressure-enhanced hybridization of the itinerant $d_{x^2-y^2}$ channel suggests increasingly efficient screening of the more localized Ni $d$ moments; we therefore examine the corresponding local spin response. Figs.~\ref{fig:susc}(a) and (b) present the static local spin susceptibility and its inverse, respectively, as functions of temperature under different pressures. In DFT+DMFT, the local spin susceptibility is obtained from the imaginary-time local spin-spin correlation function,
\[
\chi_{\mathrm{loc}}(T)=\int_0^\beta d\tau\,
\langle S_z(\tau)S_z(0)\rangle ,
\]
where $\beta=1/k_{\mathrm{B}}T$. This quantity measures the response of the local Ni $d$ spin degrees of freedom before including nonlocal magnetic correlations. Therefore, it is a useful probe of local-moment formation, orbital-dependent correlation effects, and the screening of local moments by itinerant electrons.

This quantity allows us to test directly whether the enhanced itinerancy identified above suppresses local-moment fluctuations. As shown in Fig.~\ref{fig:susc}(a), $\chi_{\mathrm{loc}}$ decreases with increasing temperature for all pressures, consistent with the presence of thermally fluctuating local moments. Meanwhile, at a fixed temperature, $\chi_{\mathrm{loc}}$ is strongly suppressed by pressure, indicating that compression weakens the local magnetic response and drives the system toward a more itinerant electronic regime. To distinguish a change in moment amplitude from a change in the screening scale, we further analyze the inverse susceptibility. Figure~\ref{fig:susc}(b) exhibits an approximately linear temperature dependence over the fitted temperature window, indicating Curie-Weiss-like behavior,
\[
\chi_{\mathrm{loc}}(T)=\frac{C}{T-\theta},
\qquad
\chi_{\mathrm{loc}}^{-1}(T)=\frac{T-\theta}{C}.
\]
Here, $C$ is the Curie constant, which is proportional to the square of the effective local moment, and $\theta$ is the Weiss temperature. The nearly pressure-independent slopes of $\chi_{\mathrm{loc}}^{-1}(T)$ suggest that the effective local moment changes only weakly with pressure. By contrast, the pressure-dependent intercept reflects a substantial change in the magnetic screening scale. Within the above convention, the extracted negative $\theta$ indicates dominant antiferromagnetic-like local spin correlations among the Ni $d$ electrons.

Since pressure mainly affects the Curie-Weiss intercept rather than the slope, we use the extracted Weiss temperature to estimate the evolution of the characteristic local screening scale. Following the standard analysis of the local spin susceptibility in correlated metals, this scale is given by
\[
T_K \simeq \frac{|\theta|}{\sqrt{2}}
\]
provides an approximate measure of the Kondo-like screening temperature~\cite{katanin2021extracting}. Here $T_K$ should not be interpreted as the Kondo temperature of an isolated impurity, but rather as a DMFT local coherence scale describing how efficiently the local Ni $d$ moments are screened by the itinerant electronic bath. Using this estimate, $T_K$ increases monotonically with pressure, rising from approximately $641$~K at 14.1~GPa to nearly four times this value at 90~GPa. This rapid increase is consistent with the pressure-enhanced hybridization and bandwidth discussed above.

Physically, the nearly occupied Ni $d_{z^2}$ bonding state retains a more localized character and contributes substantially to the local spin response, whereas the more dispersive Ni $d_{x^2-y^2}$ states provide an itinerant electronic state. Under pressure, the enhanced orbital hybridization and increased bandwidth strengthen the screening of the local moments. In the low-pressure regime, the local moments remain only partially screened, allowing strong local spin fluctuations to persist. In the high-pressure regime, the larger screening scale indicates that the local moments are more efficiently screened at temperatures well above the superconducting transition temperature.

\subsection{Interlayer Antiferromagnetic Correlations}

\begin{figure*}[htbp]
    \centering
    \includegraphics[width=0.8\textwidth]{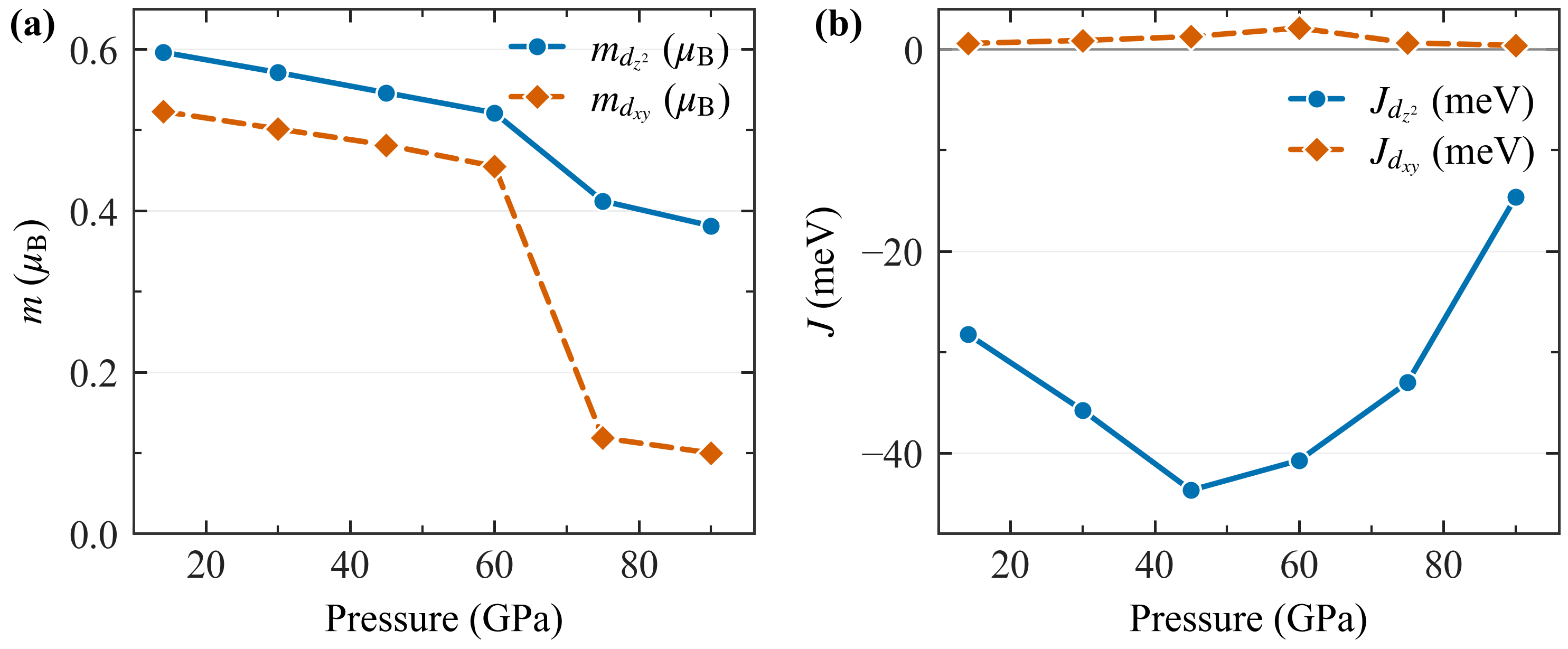}
    \caption{Pressure evolution of the orbital-resolved magnetic properties in La$_3$Ni$_2$O$_7$.
    (a) Orbital-resolved magnetic moments of the Ni $d_{z^2}$ and $d_{xy}$ orbitals obtained from spin-polarized DFT+$U$ calculations.
    (b) Orbital-resolved contributions of the Ni $d_{z^2}$ and $d_{xy}$ channels to the nearest-neighbor interlayer exchange interaction. In the present convention, negative values correspond to antiferromagnetic coupling.}
    \label{fig:mag}
\end{figure*}

The DMFT susceptibility establishes that pressure strengthens the screening of local Ni $d$ moments, but it does not directly resolve how the underlying interlayer exchange interaction evolves. To address this question independently, we performed spin-polarized DFT+$U$
calculations for the A-type antiferromagnetic state~\cite{zhu2025magnetic} and evaluated the orbital-resolved magnetic moments and exchange couplings using ABACUS combined with TB2J. In the spin-polarized solution, the significant orbital-resolved magnetic moments arise mainly from the Ni $d_{z^2}$ and $d_{xy}$ components. Therefore, the magnetic analysis in Fig.~\ref{fig:mag} focuses on these two orbital channels.

We first examine whether the pressure-enhanced itinerancy inferred from the paramagnetic calculations is also reflected in the orbital-resolved spin polarization. As shown in Figure~\ref{fig:mag}(a), the projected magnetic moments of both orbitals decrease with increasing pressure, indicating a pressure-induced crossover toward a more itinerant electronic regime. The Ni $d_{z^2}$ orbital retains a sizable moment over the whole pressure range, decreasing gradually from about $0.60~\mu_B$ at 14.1~GPa to about $0.38~\mu_B$ at 90~GPa. By
contrast, the Ni $d_{xy}$ moment is more strongly suppressed at high pressure: it decreases moderately below 60~GPa, but drops sharply above this pressure and reaches only about $0.10~\mu_B$ at 90~GPa. This behavior shows that the $d_{z^2}$ orbital carries the more robust local magnetic component, whereas the $d_{xy}$ spin polarization is more sensitive to pressure-induced itinerancy.

The reduction of the orbital moments motivates an examination of whether the interlayer magnetic coupling is suppressed in parallel.
Figure~\ref{fig:mag}(b) shows the corresponding orbital-resolved contributions to the nearest-neighbor interlayer exchange interaction. With the sign convention used here, the $d_{z^2}$ channel is antiferromagnetic, as indicated by the negative value of $J_{d_{z^2}}$. Its
magnitude is much larger than that of the $d_{xy}$ channel, demonstrating that the dominant magnetic interaction is the interlayer antiferromagnetic exchange associated with the Ni $d_{z^2}$ orbital. This coupling exhibits a nonmonotonic pressure dependence, reaching its largest magnitude of approximately $43.7$~meV around 45~GPa before being reduced at higher pressures. In contrast, the $d_{xy}$ exchange remains much smaller, with an energy scale of only a few~meV, and therefore plays a secondary role in the interlayer magnetic coupling.

Combining the moment and exchange results allows us to distinguish the pressure-enhanced bare superexchange tendency from the simultaneous
suppression of the effective magnetic response by itinerancy and screening. These results indicate that the magnetic response of pressurized La$_3$Ni$_2$O$_7$ is governed primarily by the competition between the robust $d_{z^2}$ local moment and the pressure-enhanced itinerancy. At intermediate pressures, the sizable $d_{z^2}$ moment and strong interlayer antiferromagnetic exchange provide favorable conditions for pronounced spin fluctuations. At higher pressures, however, the local moments are substantially quenched, especially in the $d_{xy}$ channel, while the $d_{z^2}$ exchange is also reduced from its maximum value. The system therefore evolves toward a more itinerant magnetic regime in which local spin fluctuations are weakened. Combining these results with the calculated pressure evolution of electron itinerancy provides the following picture. At low pressures, the antiferromagnetic exchange between the local moments associated with the $d_{z^2}$ orbitals is governed primarily by superexchange and is enhanced by the increasing hopping amplitude $t_{d_{z^2}}$. At intermediate and high pressures, the increasingly itinerant $d_{x^2-y^2}$ electrons more effectively screen the local moments, leading to a pronounced reduction of the antiferromagnetic exchange.

\section{Discussion}
The experimentally observed monotonic suppression of the superconducting transition temperature ($T_c$) under high pressure in La$_3$Ni$_2$O$_7$ presents a compelling challenge to conventional pairing paradigms. To elucidate the microscopic origin of this behavior, we examine the intricate interplay between orbital-selective electronic correlations and inter-orbital hybridization. Our first-principles calculations reveal that applied pressure systematically enhances the interlayer hopping integral ($t$) associated with the Ni $d_{z^2}$ orbitals. Consequently, the bare superexchange interaction, scaling as $4t^2/U$, exhibits a monotonic increase with pressure. In a purely localized scenario, this would naturally elevate the pairing strength and $T_c$. However, the actual high-pressure behavior is governed by the multi-orbital nature of the system, specifically the orbital-selective Mott physics inherent to the $d_{z^2}$ and $d_{x^2-y^2}$ manifolds.

Within the orbital-selective framework, the $d_{z^2}$ orbitals maintain a relatively localized character, providing the necessary local magnetic moments for spin-fluctuation-mediated pairing, while the $d_{x^2-y^2}$ orbitals are inherently more itinerant. Our DMFT calculations demonstrate that pressure drives the $d_{x^2-y^2}$ states further into the itinerant regime, significantly enhancing their delocalization. Crucially, this pressure-induced itinerancy is accompanied by a strengthened hybridization between the localized $d_{z^2}$ and itinerant $d_{x^2-y^2}$ orbitals. As a direct consequence, the Kondo screening of the $d_{z^2}$ local moments by the $d_{x^2-y^2}$ conduction electrons is dramatically amplified, as evidenced by the significant monotonic rise of the Kondo temperature ($T_K$) with increasing pressure.

The macroscopic superconducting properties are thus determined by the fierce competition between the pressure-enhanced superexchange ($4t^2/U$) and the concurrently strengthened Kondo screening. In the high-pressure regime investigated here, the rapid escalation of $T_K$ dominates the low-energy physics. The itinerant $d_{x^2-y^2}$ electrons effectively ``wash out'' the localized $d_{z^2}$ magnetic moments, leading to a substantial reduction in the effective magnetic exchange coupling ($J_{\text{eff}}$). Since $J_{\text{eff}}$ serves as the fundamental energy scale for the spin-fluctuation pairing glue in this system, its suppression directly undermines the superconducting instability. Therefore, the monotonic decrease of $T_c$ under high pressure is not merely a structural effect, but rather a manifestation of the system traversing beyond the optimal quantum critical balance. The excessive Kondo screening ultimately quenches the very local moments required for unconventional superconductivity, providing a coherent microscopic explanation for the high-pressure dome-shaped phase diagram observed in La$_3$Ni$_2$O$_7$. This work sheds new light on the superconducting mechanisms of nickelates and offers crucial guidance for future explorations.

\section{Conclusion}

In summary, our DFT, cRPA, DFT+DMFT, and TB2J calculations reveal that the pressure-induced suppression of superconductivity in bilayer La$_3$Ni$_2$O$_7$ arises from a competition between enhanced itinerancy and magnetic exchange. Pressure increases the interlayer $d_{z^2}$ hopping and bare superexchange scale $4t^2/U$, while continuously reducing $U/W$. However, the dominant interlayer antiferromagnetic $d_{z^2}$ exchange is nonmonotonic, peaking at intermediate pressures and weakening at higher pressures due to pressure-induced moment reduction and itinerancy. Concurrently, the $d_{x^2-y^2}$ orbital becomes increasingly itinerant, enhancing the screening of local Ni moments, as reflected in the rising Kondo scale $T_K$. Thus, superconductivity is favored at intermediate pressures where exchange remains strong but screening is not yet dominant. At higher pressures, enhanced itinerancy and screening suppress the magnetic response and hence $T_c$. Our findings pave the way for future investigations into the pressure-dependent pairing mechanism and the role of orbital selectivity in nickelate superconductors.

\begin{acknowledgments}
This work is supported by the National Key Research and Development Program of China (under Grant No.~2024YFA1408601), the National Natural Science Foundation of China (under Grant No.~12474241, No.~12404188), and the Presidential Foundation of CAEP (under Grant No.~YZJJZQ2024014).
\end{acknowledgments}

\appendix
\section{Additional DFT and DMFT Results}
\label{sec:appendix}

\begin{figure*}[h!]
    \centering
    \includegraphics[width=0.95\textwidth]{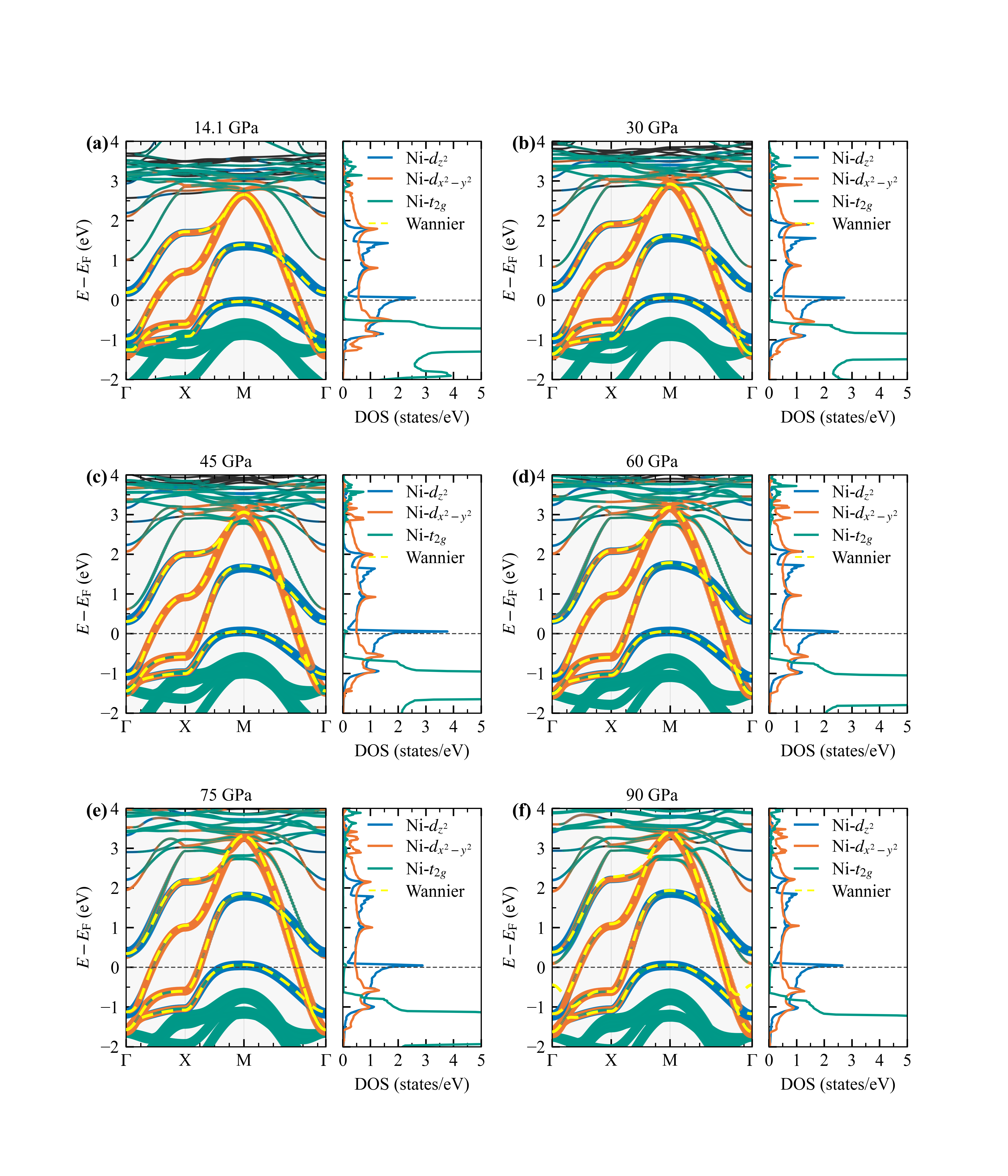}
    \caption{Orbital-resolved DFT electronic structures of La$_3$Ni$_2$O$_7$ under compression at 80~K. The solid lines indicate the DFT bands projected onto the Ni $d_{z^2}$, $d_{x^2-y^2}$, and $t_{2g}$ orbitals, while the yellow dashed lines show the corresponding Wannier-interpolated bands. The orbital-resolved densities of states are displayed alongside the band structures. Panels (a)--(f) correspond to 14.1, 30, 45, 60, 75, and 90~GPa, respectively. The Fermi level is set to zero.} \label{fig:appendix1}
\end{figure*}

Figure~\ref{fig:appendix1} complements the representative 14.1~GPa results shown in Fig.~\ref{structure} by displaying the complete pressure evolution of the DFT electronic structure. At all investigated pressures, the states near the Fermi level remain dominated by the Ni
$e_g$ orbitals, whereas the nearly filled $t_{2g}$ manifold lies mainly at lower energies. The $d_{x^2-y^2}$ bands retain a comparatively strong in-plane dispersion, while the bonding $d_{z^2}$ branch remains less dispersive. With increasing pressure, the overall $e_g$ bandwidth grows and the low-energy bands undergo a continuous reconstruction, consistent with the pressure-enhanced hopping and reduced $U/W$ discussed in the main text. The close agreement between the DFT bands and the Wannier-interpolated bands over the entire pressure range also confirms that the same low-energy Wannier subspace provides a reliable basis for the pressure-dependent cRPA analysis.

\begin{figure*}[h!]
    \centering
    \includegraphics[width=0.95\textwidth]{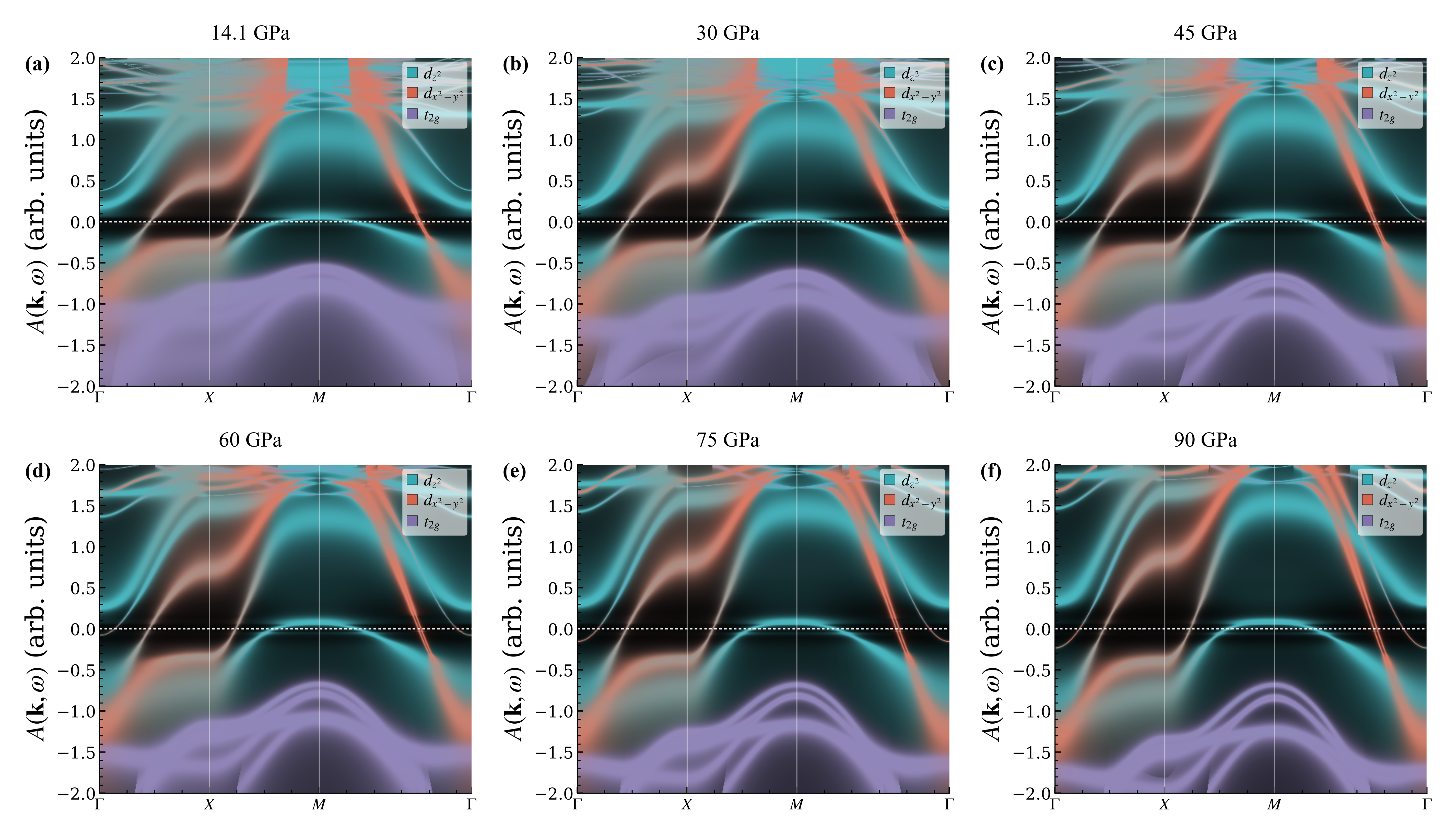}
    \caption{DFT+DMFT momentum-resolved spectral functions $A(\mathbf{k},\omega)$ of La$_3$Ni$_2$O$_7$ at 80~K.
    The spectral weight is resolved into the Ni $d_{z^2}$, $d_{x^2-y^2}$, and $t_{2g}$ orbital contributions.
    Panels (a)--(f) correspond to 14.1, 30, 45, 60, 75, and 90~GPa, respectively. The horizontal dashed line denotes the Fermi level.} \label{fig:appendix2}
\end{figure*}

The complete set of momentum-resolved spectra in Figure~\ref{fig:appendix2} confirms that the orbital-selective renormalization discussed in Fig.~\ref{fig:band_dmft} persists throughout the investigated pressure range. The low-energy $d_{z^2}$-derived branch near the $M$ point remains more strongly renormalized than the more dispersive $d_{x^2-y^2}$ bands. Pressure continuously shifts and reshapes these quasiparticle bands near the Fermi level, producing the gradual Fermi-surface reconstruction described in the main text, including the increasing visibility of the $d_{z^2}$-derived $\gamma$ pocket and the high-pressure appearance of the additional $\delta$ pocket. Meanwhile, the $t_{2g}$ spectral weight remains concentrated predominantly below the Fermi level and contributes only weakly to the low-energy quasiparticle states.

\begin{figure*}[h!]
    \centering
    \includegraphics[width=0.95\textwidth]{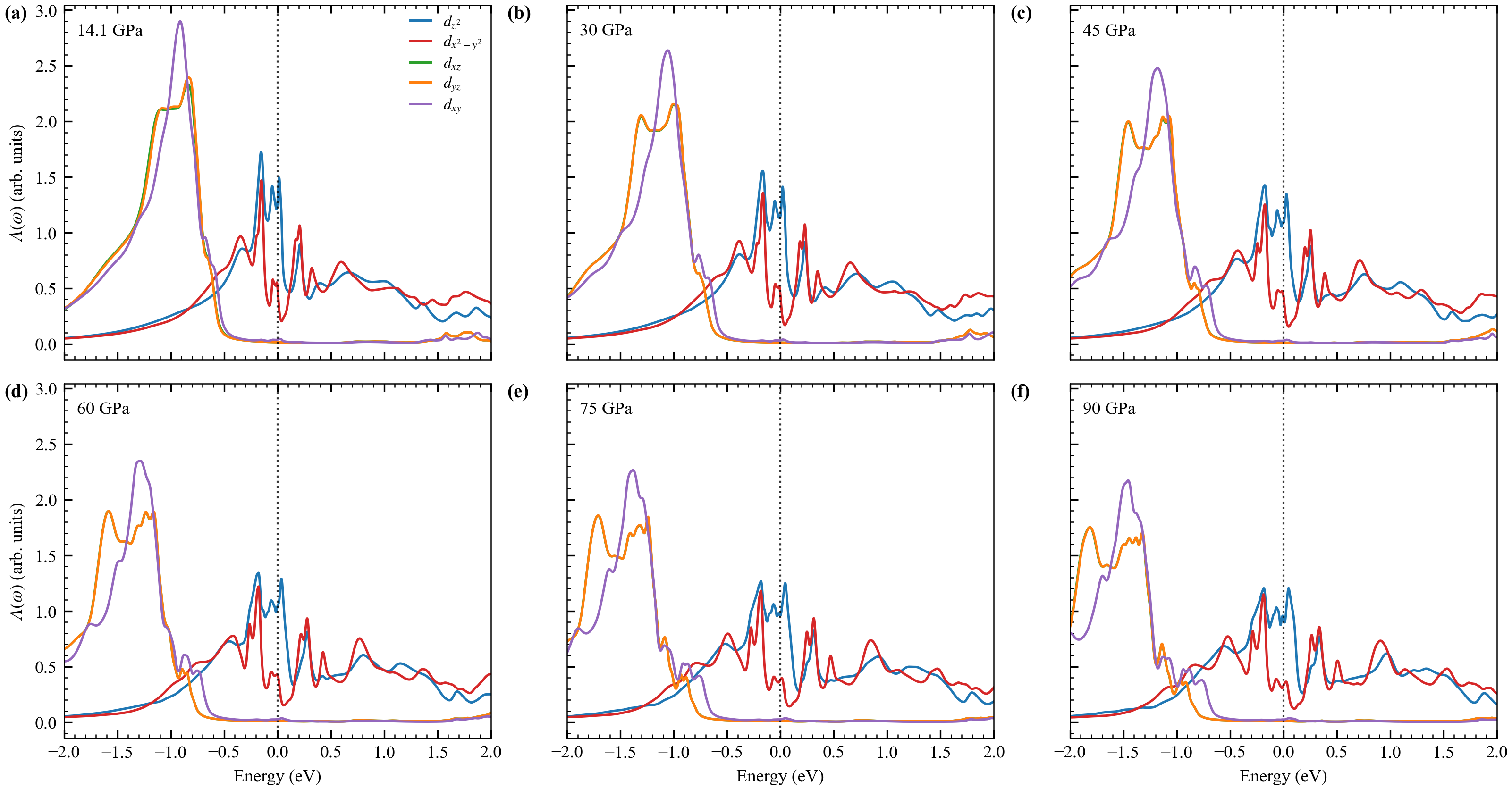}
    \caption{DFT+DMFT orbital-resolved local spectral functions $A(\omega)$ of La$_3$Ni$_2$O$_7$ at 80~K.
    Panels (a)--(f) correspond to 14.1, 30, 45, 60, 75, and 90~GPa, respectively. The vertical dotted line marks the Fermi level.}
    \label{fig:appendix3}
\end{figure*}

Figure~\ref{fig:appendix3} provides the corresponding local view of the orbital-resolved spectral-weight redistribution. Finite spectral weight from both $d_{z^2}$ and $d_{x^2-y^2}$ orbitals is present around the Fermi level at all pressures, confirming the multiorbital metallic character of pressurized La$_3$Ni$_2$O$_7$. The sharper low-energy structures of these two $e_g$ orbitals contrast with the predominantly occupied $t_{2g}$ weight at lower energies. As pressure increases, the $e_g$ spectral features broaden and redistribute continuously rather than developing a correlation-induced gap. This evolution is consistent with enhanced orbital hybridization and a progressive reduction of the relative correlation strength, while the distinct line shapes of the two $e_g$ components retain the orbital-selective character emphasized in the main text.

\begin{figure*}[h!]
    \centering
    \includegraphics[width=0.95\textwidth]{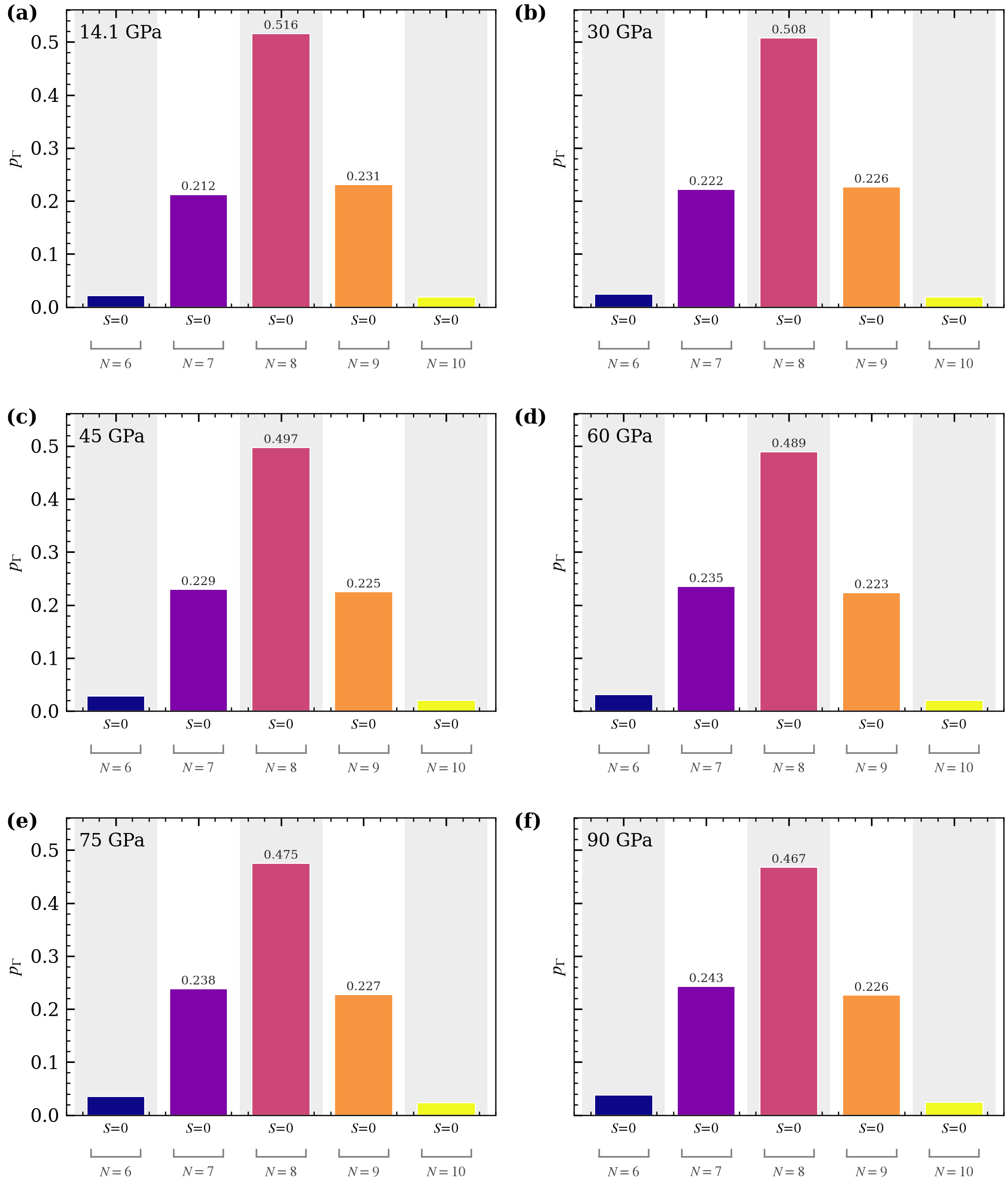}
    \caption{DMFT valence-state histograms of La$_3$Ni$_2$O$_7$ at different pressures at 80~K. The probabilities $p_{\Gamma}$ of the atomic configurations are denoted by using good quantum numbers $N$ (total Ni $3d$ occupancy). Panels (a)--(f) correspond to 14.1, 30, 45, 60, 75, and 90~GPa, respectively.} \label{fig:appendix4}
\end{figure*}

The valence-state histograms in Fig.~\ref{fig:appendix4} show that the DMFT impurity is characterized by substantial valence state fluctuations rather than by a single integer Ni $3d$ configuration. The $N=8$ sector has the leading statistical weight over the entire pressure range, but its probability decreases gradually from 0.516 at 14.1~GPa to 0.467 at 90~GPa. At the same time, the $N=7$ contribution increases from 0.212 to 0.243, whereas the $N=9$ weight remains close to 0.22$\sim$0.23 and the $N=6$ and $N=10$ sectors remain minor. This redistribution indicates that pressure enhances charge fluctuations among neighboring valence sectors, consistent with the increasingly itinerant and strongly hybridized electronic environment inferred from the band and hybridizatio function results.

\begin{figure*}[h!]
    \centering
    \includegraphics[width=0.8\textwidth]{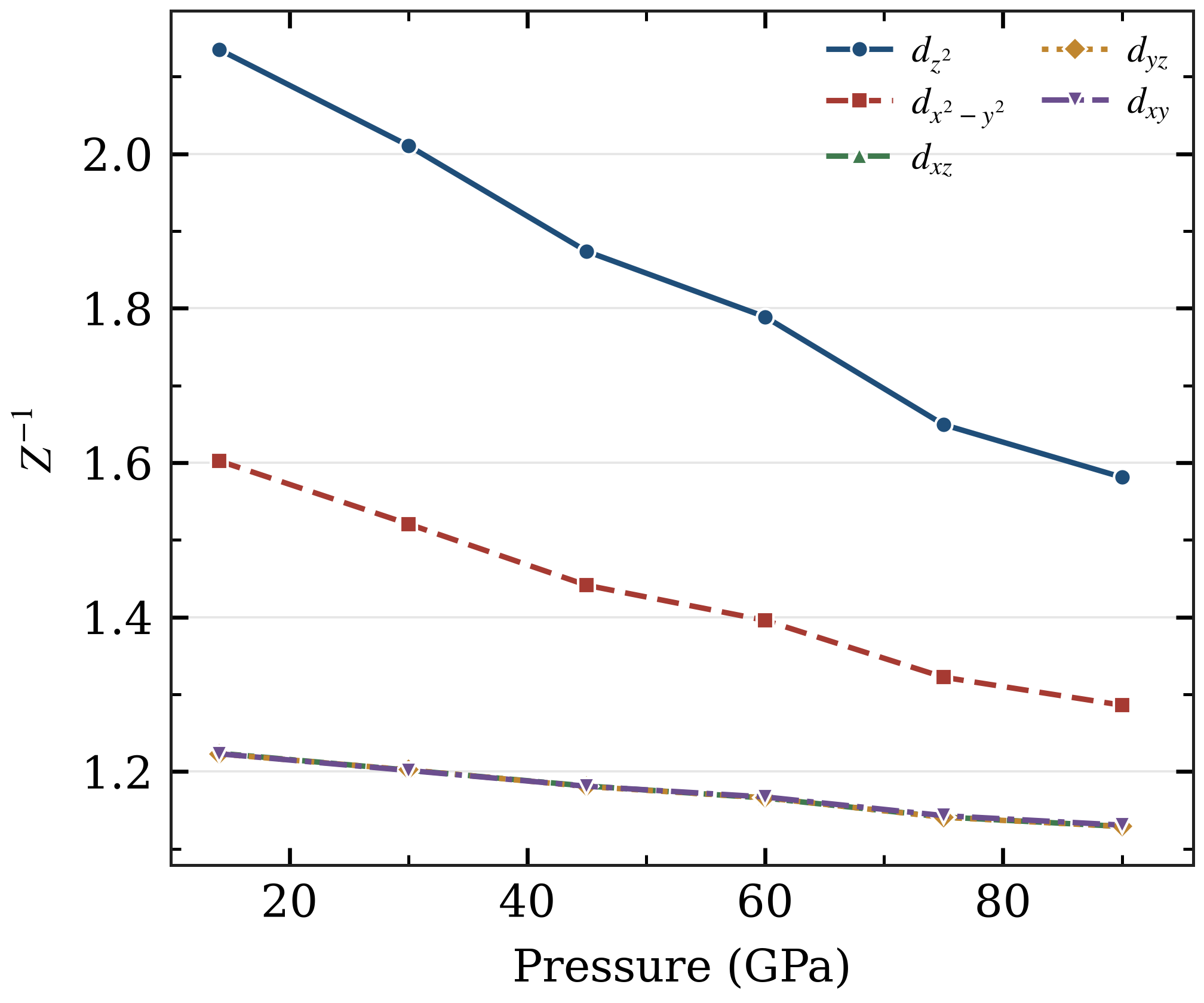}
    \caption{Pressure dependence of the inverse quasiparticle renormalization factors $Z^{-1}$ for the Ni $3d$ orbitals in
    La$_3$Ni$_2$O$_7$ at 80~K. Larger values of $Z^{-1}$ indicate stronger quasiparticle mass enhancement.} \label{fig:appendix5}
\end{figure*}

Figure~\ref{fig:appendix5} quantifies the pressure-dependent reduction of the orbital-resolved correlation strength. The $d_{z^2}$ orbital exhibits the largest inverse quasiparticle weight throughout the pressure range, with $Z^{-1}$ decreasing from approximately 2.13 at 14.1~GPa to 1.58 at 90~GPa. The $d_{x^2-y^2}$ value decreases from about 1.60 to 1.29, while
the three $t_{2g}$ orbitals remain only weakly renormalized, with $Z^{-1}$ close to 1.1$\sim$1.2. Thus, all Ni $d$ orbitals become progressively more itinerant under compression, but a clear orbital selectivity remains: the $d_{z^2}$ states are the most strongly renormalized, followed by $d_{x^2-y^2}$, whereas the nearly filled $t_{2g}$ states are comparatively weakly correlated. This trend supports the main-text interpretation that pressure reduces the overall relative correlation strength while preserving a more localized $d_{z^2}$ component.

\end{document}